\documentclass[reprint,amsmath,amssymb,aps]{revtex4-2}

\usepackage{graphicx}
\usepackage{dcolumn}
\usepackage{bm}

\begin{document}
	
	\preprint{APS/123-QED}
	
	\title{Test of Artificial Neural Networks in Likelihood-free Cosmological Constraints: A Comparison of IMNN and DAE}
	
	\author{Jie-Feng Chen}
	\affiliation{Institute for Frontiers in Astronomy and Astrophysics,Department of Astronomy,Beijing Normal University, Beijing 102206, China.}
	\affiliation{Department of Astronomy, Beijing Normal University, Beijing 100875, China}
	
	\author{Yu-Chen Wang}%
	\affiliation{
	Kavli Institute for Astronomy and Astrophysics, Peking University, Beijing 100871, China
	}%
	\affiliation{
	Department of Astronomy, School of Physics, Peking University, Beijing 100871, China
	}%
	\affiliation{%
	Department of Physics, Beijing Normal University, Beijing 100875, China
	}%
	
	\author{Tingting Zhang}
	\email{101101964@seu.edu.cn}
	\affiliation{
		College of Command and Control Engineering, Army Engineering University, Nanjing 210017, China
	}%
	
	\author{Tong-Jie Zhang}
	\email{tjzhang@bnu.edu.cn}
	\affiliation{Institute for Frontiers in Astronomy and Astrophysics,Department of Astronomy,Beijing Normal University, Beijing 102206, China.}
	\affiliation{%
		Department of Astronomy, Beĳing Normal University, Beĳing 100875, China}%
	\affiliation{%
		Institute for Astronomical Science, Dezhou University,
		Dezhou 253023, China}%
	\date{\today}
	
	\begin{abstract}
	In the procedure of constraining the cosmological parameters with the observational Hubble data, the combination of Masked Autoregressive Flow and Denoising Autoencoder can perform a good result. The above combination extracts the features from OHD with DAE, and estimates the posterior distribution of $H_0,\, \Omega_{m},\, \Omega_\Lambda$ with MAF. We ask whether we can find a better tool to compress large data in order to gain better results while constraining the cosmological parameters. Information maximising neural networks, a kind of simulation-based machine learning technique, was proposed at an earlier time. In a series of numerical examples, the results show that IMNN can find optimal, non-linear summaries robustly. In this work, we mainly compare the dimensionality reduction capabilities of IMNN and DAE. We use IMNN and DAE to compress the data into different dimensions and set different learning rates for MAF to calculate the posterior. Meanwhile, the training data and mock OHD are generated with a simple Gaussian likelihood, the spatially flat $\Lambda\mathrm{CDM}$ model and the real OHD data. To avoid the complex calculation in comparing the posterior directly, we set different criteria to compare IMNN and DAE.

	\end{abstract}
	
	\maketitle
	
	
	\section{Introduction}
	
	Constraining cosmological parameters is a basic task in cosmology. To evaluate the parameters, the common method is to calculate an intractable likelihood directly to perform Bayesian inference with the existing observational datasets, e.g.  observational Hubble parameter data (OHD, \citet{2017Bayesian}), type Ia supernovae (SNe Ia, \cite{2017The}), cosmic microwave background \cite{2017The}, and large-scale structures \cite{2012Measurements} \cite{pan2020cosmological}. Approximate Bayesian Computation(ABC) has also shown good performance in many astronomical tasks, such as galaxy evolution \cite{10.1111/j.1365-2966.2012.21371.x} and SN Ia cosmology \cite{Weyant_2013}. Nevertheless, according to \cite{2016Fast}, conventional ABC algorithms may suffer from noisy computations.
	
	In the past few decades, the artificial neural networks (ANN) developed rapidly and were gradually used to constrain the cosmological parameters \cite{2022Estimating} \cite{2022Constraining} \cite{2019Parameters} \cite{2020Constraining}. Recently, a likelihood-free inference procedure using Denoising autoencoder (DAE) and Masked Autoregressive Flow (MAF) was proposed by us \cite{2021Likelihood}. In our previous work, the combination of MAF and DAE was compared to MCMC, which is the Markov Chain Monte Carlo method, and behaved well in calculating the posterior distribution ($P(\boldsymbol{\theta}|\textbf{\emph{H}}_{\mathrm{obs}})$) of $\Omega_\Lambda,\Omega_m,$ and $H_0$. We proved that MAF could give similar results as MCMC, which means that at least MAF could be the substitute while we estimate the cosmological parameters. MAF was proposed by \cite{2017arXiv170507057P}, in their several experiments, MAF gave accurate estimations of distributions and did well in likelihood-free inference \cite{papamakarios2019sequential}. DAE \cite{10.1145/1390156.1390294} is an ANN that can encode data by extracting the data features. With the DAE, we can obtain low-dimensional representative features of the input data without an artificial choice of statistics.
	
	The higher-dimensional data from simulation or computational resources is inevitable in likelihood-free inference. For this reason, we need to find a good tool to reduce the dimensionality of data. At an earlier time, \cite{2018Automatic} proposed a kind of ANN named "information maximizing neural networks" (IMNNs), which can transform data into summaries by maximizing the Fisher information at fiducial values. In the examples proposed by \cite{2018Automatic} and \cite{10.1093/mnras/stz1960}, IMNN performed well in finding the informative data summaries, which means maybe we can test whether IMNN can be a substitute for DAE.
	
	In this work, we attempt to compare the dimensionality reduction capabilities of IMNN and DAE. Like the procedure of constraining cosmological parameters applied by \cite{2021Likelihood}, we use DAE and IMNN to reduce the higher-dimensional OHD data and then use MAF to estimate the distributions of cosmological parameters with the low-dimensional features. Besides, we also estimate the distribution, which will be treated as the standard distribution, with MAF and the original-dimensional OHD data. In the rest of this article, we use MAF-IMNN to represent the combination of MAF and IMNN, and MAF-DAE to represent the combination of MAF and DAE. In section \ref{sec:s2}, we review the procedure of cosmological constraints using MAF-DAE. In section \ref{sec:s3}, we discuss the theory of IMNN. In section \ref{sec:s4}, we will show the results of constraint with OHD in different ways, and explore the possibility of evaluating parameters with MAF-IMNN. In section \ref{sec:s5}, we compare the DAE and IMNN with different criteria. Finally, in Section \ref{sec:s6}, we conclude and discuss.
	
	\section{MAF-DAE for Parameter Constraint}\label{sec:s2}
	
	MAF, the combination of normalizing flow and Masked Autoencoder for Distribution Estimation (MADE \citep{2015MADE}, one kind of autoregressive model), was proposed by \cite{2017arXiv170507057P}. MADE and normalizing flows are two kinds of neural density estimators, which can estimate the density distribution of the parameters.
	
	\begin{figure}[h]
		\centering                                      
		\includegraphics[width=0.5\textwidth]{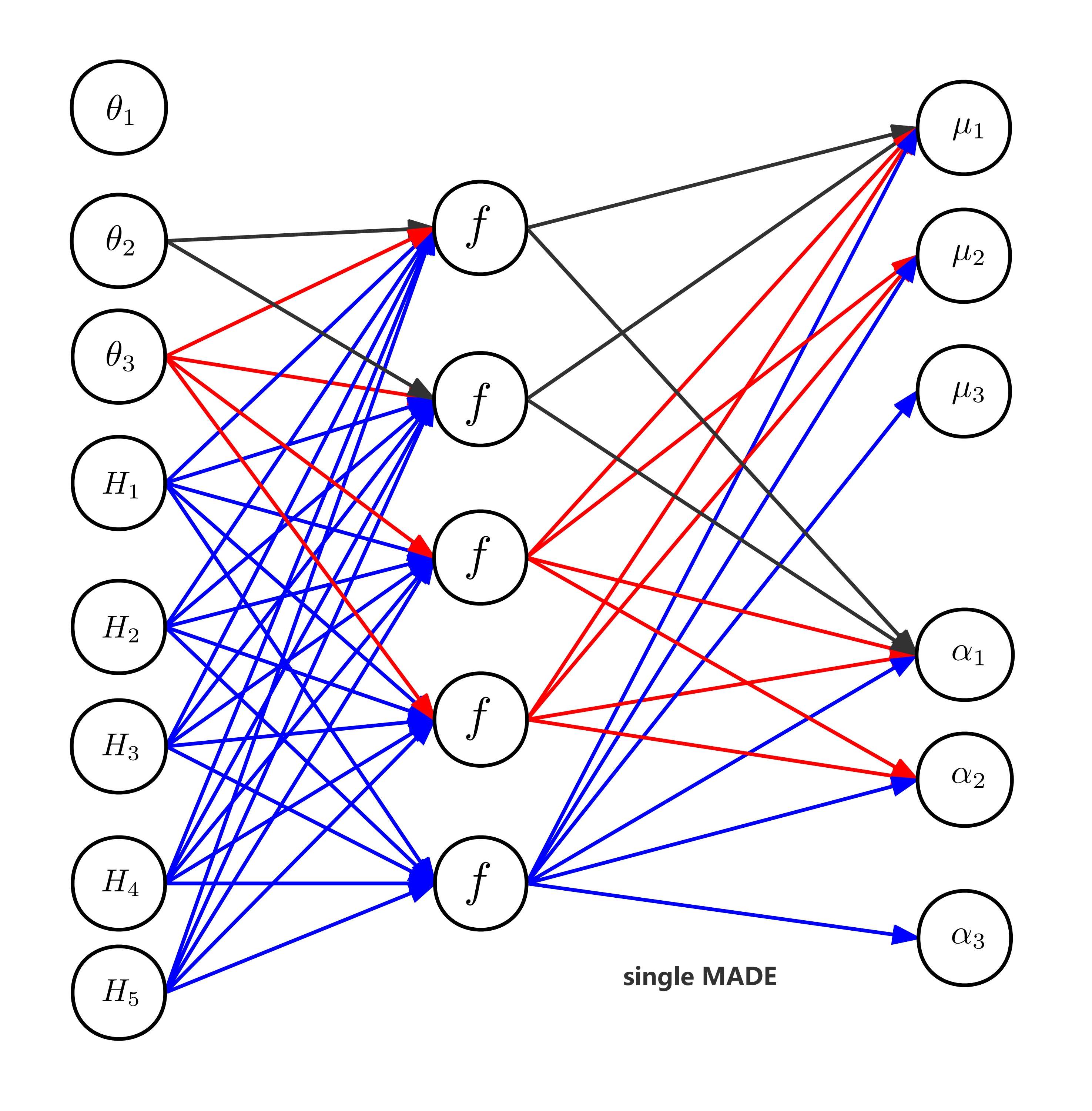}    
		\caption{The concise structure of the MADE. The blue lines means full connection. The black lines and the red lines mean a set of connections is removed ensuring that MADE satisfies the autoregressive property. In autoregressive property, a unit can only connect to one more advanced unit. To explain the MAF, we only draw 5 $H_i$s (the OHD) correlating to 3 $\theta_i$s ($\Omega_{\Lambda}$,$\Omega_{m}$ and $H_0$) in this figure, but in this work we actually applied 31 $H_i$s. }                            
		\label{fig:maf}                                     
	\end{figure}
	
	With the MADE, the conditional distribution $P(\textbf{\emph{x}}|\textbf{\emph{y}})$ can be written in the form:
	\begin{equation}\label{eq1}
	P(\textbf{\emph{x}}|\textbf{\emph{y}}) = \prod_d P(x_d|\boldsymbol{x}_{1:d-1},y),
	\end{equation}
	where $\boldsymbol{x}_{1:d-1}=(x_1,x_2,...,x_{d-1})$, which means $\textbf{\emph{x}}$ has $\textbf{\emph{d}}$ dimensions. And then Eq. (\ref{eq1}) will be parameterized into Gaussian distribution, the mean and the log standard deviation will be calculated by the neural network.  In other words, we can obtain the parameters of all of these conditional distributions. The concise structure of the MADE is shown in Fig. \ref{fig:maf}.
	
	According to the normalizing flows \cite{2015Variational}, the density $P(x)$ can be obtained from a base density $\pi_u(\boldsymbol{u})$ with an invertible differentiable transformation $f$:
	\begin{equation}
	P(x) = \pi_u(f^{-1}(x))|\mathrm{det}(\frac{\partial f^{-1}}{\partial x})|,
	\end{equation}
	where $\boldsymbol{u} = f^{-1}(x)$ and $\boldsymbol{u} \sim \pi_u(\boldsymbol{u})$ (usually $ \pi_u(\boldsymbol{u})$ is a standard Gaussian distribution ($\boldsymbol{u}\sim\mathcal{N}(0,\mathrm{I})$)). With the normalizing flows and autoregressive models, each of the conditionals $P(x_d|\boldsymbol{x}_{1:d-1},\textbf{\emph{y}})$ can be parameterized as Gaussian distribution. In this case, the $d^{th}$ conditional is
	\begin{equation}
	P(x_d|x_{1:d-1}) = \mathcal{N}(x_d|\mu_d,(\exp\,\, \alpha_d)^2),
	\end{equation}
	\begin{equation}
	\pi(u_d)=\mathcal{N}(u_d;0,1),
	\end{equation}
	and
	\begin{equation}
	x_d=f_d(u_d;\alpha_d,\mu_d)=\exp(\alpha_d)\,u_d+\mu_d.
	\end{equation}
	The unconstrained scalar functions $\mu_i=f_{\mu i}(x_{1:i-1})$ and $\alpha_i=f_{\alpha i}(x_{1:i-1})$ compute the mean and log standard respectively.
	
	When doing the cosmological inference, we can represent $\boldsymbol{x}$ as the $\boldsymbol{\theta}$ in the Hubble parameters such as $\Omega_\Lambda,\Omega_m,H_0$, and represent $\textbf{\emph{y}}$ as $\textbf{\emph{H}}_{\mathrm{obs}}$. In this case $\boldsymbol{\theta}_i$ can be written in the form:
	\begin{equation}
	\theta_i=u_i \exp(\alpha_i)+\mu_i,
	\end{equation}
	\begin{equation}
	P(\theta) = \pi_u(f^{-1}(\theta))|det(\frac{\partial f^{-1}(\theta)}{\partial \theta})|,
	\end{equation}
	where $\mu_i=f_{\mu i}(\boldsymbol{\theta}_{1:i-1}),\,\,\alpha_i=f_{\alpha i}(\boldsymbol{\theta}_{1:i-1})$ and $u_i\sim\mathcal{N}(0,1)$, so with the MADE we can get the $\boldsymbol{\theta}=f(\boldsymbol{u};\textbf{\emph{H}})$ where $\boldsymbol{u}\sim\mathcal{N}(0,I)$. A single MADE may not fit the distribution well, which means that the corresponding random numbers $\boldsymbol{u}=f^{-1}(\boldsymbol{\theta};\textbf{\emph{H}})$ transformed from the training data $\boldsymbol{\theta}$  were not standard Gaussian (Also $\boldsymbol{\theta}$ = $f(\boldsymbol{u};\textbf{\emph{H}}$)). To improve the performance of MADE, we can stack several MADEs as a normalizing flow. According to \cite{2017arXiv170507057P},  Masked Autoregressive Flow (MAF) is the implementation of stacking MADEs into a flow. The loss function of MAF is defined by the negative log probability:
	\begin{equation}
	L=-\sum_n \mathrm{ln}P(\boldsymbol{\theta}_n|\textbf{\emph{H}}_n).
	\end{equation}
	where $n$ means the $n^{th}$ data in the training data.
	
	The training set of the MAF should be the $\{\boldsymbol{\theta}_n,\textbf{\emph{H}}_n\}$, where the $\boldsymbol{\theta}$ means $\Omega_{\Lambda}$, $\Omega_{m}$ and $H_0$, and $\textbf{\emph{H}}$ means different dimensional mock OHD. In this work, we trained MAF to find the correlation between $\boldsymbol{\theta}$ and $\textbf{\emph{H}}$ (5-dimensional $H$, 10-dimensional $H$, 15-dimensional $H$, 20-dimensional $H$, 31-dimensional $H$). After training, MAF can be used to estimate the $P(\boldsymbol{\theta}|\textbf{\emph{H}}_{\mathrm{obs}})$ with $H_{\mathrm{obs}}$, which is in its own 31-dimension or being compressed into 20-dimension, 15-dimension, 10-dimension and 5-dimension. The input of a trained MAF is a set of $H_{\mathrm{obs}}$, while the output is 100000 (we can also set another number such as 10000, 50000.) sets of $\Omega_{\Lambda},\Omega_{m}$ and $H_0$, which can be used to calculated the $P(\boldsymbol{\theta}|\textbf{\emph{H}}_{\mathrm{obs}})$ directly. Certainly, we can also input $H_{\mathrm{fid}}$ and calculate the $P(\boldsymbol{\theta}|\textbf{\emph{H}}_{\mathrm{fid}})$.
	
	\begin{figure}[h]
		\centering                                      
		\includegraphics[width=0.45\textwidth]{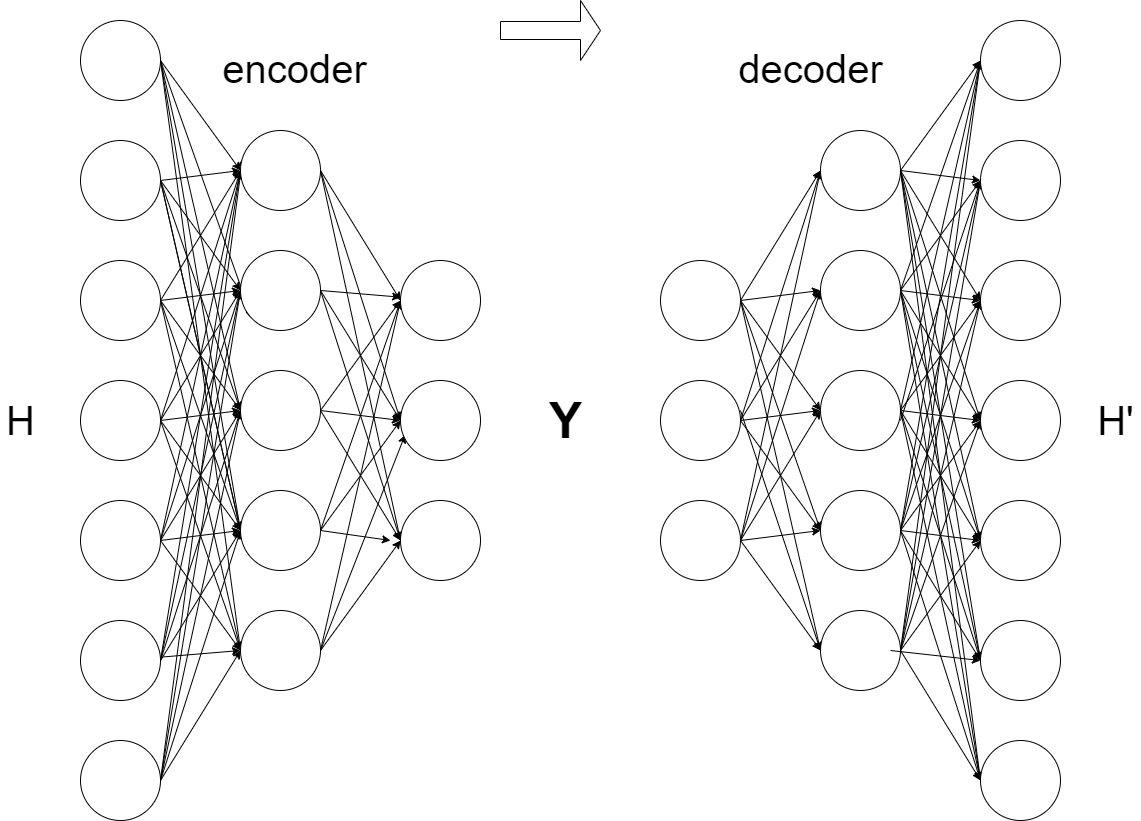}    
		\caption{The concise structure of the autoencoder and the procedure of compressing data. An autoencoder consists of an encoder and a decoder. With the encoder, the input $\textbf{\emph{H}}$ can be compressed into lower-dimensional $\textbf{\emph{y}}$, so the information in $\textbf{\emph{H}}$ can be represented by $\textbf{\emph{y}}$. With the decoder, $\textbf{\emph{y}}$ can be reconstructed to original-dimensional $\textbf{\emph{H}}'$. Usually, we train the autoencoder by minimizing the error between $\textbf{\emph{H}}$ and $\textbf{\emph{H}}'$.}
		\label{fig:dae}                                     
	\end{figure}

	\subsection{Denoising autoencoders (DAE)} 
	
	DAE is a special kind of autoencoder. A basic autoencoder is in a special neural network architecture, which is composed of an encoder and a decoder, can learn efficient, lower-dimensional codings of the input data. The autoencoder is trained with unsupervised learning to obtain lower-dimensional features of the input data by encoder. In the output part of the autoencoder (decoder), the lower-dimensional features can be reconstructed to original-dimensional data. Therefore, the input layer has the same number of neurons as the output layer. The training of the autoencoder is to minimize the error between the input and the output. The concise structure of the autoencoder is shown in Fig. \ref{fig:dae}. 
	
	DAE is trained with noise-free reconstruction criterion and noisy inputs, so that it can not only extract the robust features from the input data but also significantly reduce the noise. In this work, the DAE was trained with noise-free fiducial values $\textbf{\emph{H}}_{\mathrm{fid}}$ as labels and noisy simulated data $\textbf{\emph{H}}$ as the inputs in order to reduce the noise level of the $\textbf{\emph{H}}_{\mathrm{obs}}$ and preserve more information. After training, DAE can compress $\textbf{\emph{H}}_{\mathrm{obs}}$ to low-dimensional $\textbf{\emph{y}}$ ($\textbf{\emph{y}} = f_e(\textbf{\emph{H}})$). Usually, an autoencoder is trained by minimizing the reconstruction error, i.e. the mean squared error (MSE) between reconstructed data $\textbf{\emph{H}}'$ and the label $\textbf{\emph{H}}_{\mathrm{fid}}$. However, to make sure $\textbf{\emph{y}}$ may contain more information about $\boldsymbol{\theta}$ and avoid giving too big variance of $P(\textbf{\emph{y}}|\boldsymbol{\theta})$, our previous work\cite{2021Likelihood} proposed a complete batch loss function to require the mean of the conditional $P(\textbf{\emph{y}}|\boldsymbol{\theta})$ relies linearly on $\boldsymbol{\theta}$. The loss function consists of reconstruction MSE and encoding variance:
	\begin{equation}\label{eqtheta}
	\begin{split}
	L_{AE}=\mathrm{mean}\{(\textbf{\emph{X}}'-\textbf{\emph{X}}_{\mathrm{fid}})\circ(\textbf{\emph{X}}'-\textbf{\emph{X}}_{\mathrm{fid}})\} \\+\mathrm{var}\{\textbf{\emph{Y}}-\boldsymbol{\Theta}\boldsymbol{\Theta}^+\textbf{\emph{Y}}\},
	\end{split}
	\end{equation}
	where 
	\begin{equation}
	\textbf{\emph{X}}_{\mathrm{fid}}=\left(\begin{array}{c} 
	\textbf{\emph{H}}^T_{\mathrm{fid},1}  \\ 
	\textbf{\emph{H}}^T_{\mathrm{fid},2}  \\ 
	. \\ 
	. \\ 
	. \\ 
	\end{array}\right)
	,\textbf{\emph{X}}_{\mathrm{fid}}=\left(\begin{array}{c} 
	\textbf{\emph{H}}'^T_{1}  \\ 
	\textbf{\emph{H}}'^T_{2}  \\ 
	\vdots \\ 
	\vdots \\ 
	\vdots \\ 
	\end{array}\right)
	\end{equation}
	and 
	\begin{equation}
	\textbf{\emph{Y}}=\left(\begin{array}{c} 
	\textbf{\emph{y}}^T_{1}  \\ 
	\textbf{\emph{y}}^T_{2}  \\ 
	\vdots \\ 
	\vdots \\ 
	\vdots \\ 
	\end{array}\right)
	,\boldsymbol{\Theta}=\left(\begin{array}{c} 
	1\,\, \boldsymbol{\theta}^T_{1}  \\ 
	1\,\, \boldsymbol{\theta}^T_{2}  \\ 
	\vdots \\ 
	\vdots \\ 
	\vdots \\ 
	\end{array}\right).
	\end{equation}
	The $\boldsymbol{\Theta}^+$ in Eq. (\ref{eqtheta}) is the pseudoinverse (Moore-Penrose inverse) of $\boldsymbol{\Theta}$. In this way, the loss function can be easily evaluated on the training set. In this work, we stick with this training method.
	
	\subsection{The simulated data} 
	
\begin{figure}[t]
		\centering                                      
		\includegraphics[width=0.55\textwidth]{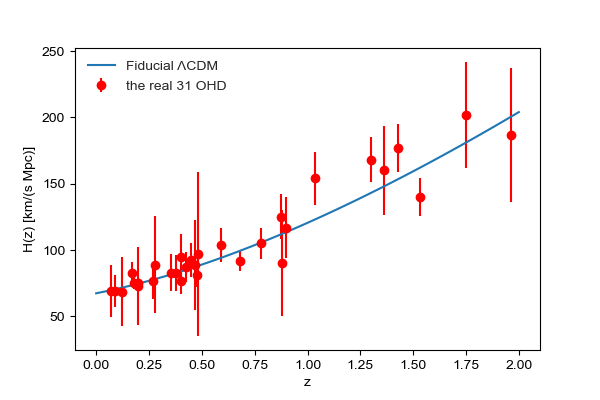}    
		\caption{The 31 real OHD datapoints and spatially flat $\Lambda\mathrm{CDM}$ model.}
		\label{fig:realohd}                                     
	\end{figure}

	The real OHD is composed of $z_i,H(z_i)$ and $\sigma_i $, where $z_i$ is the redshift, and $H(z_i)$ is the corresponding Hubble parameter and $\sigma_i$ is the corresponding uncertainty. The 31 OHD data we use in this work are evaluated with the cosmic chronometer method, which are given in \cite{Jimenez_2003}, \cite{PhysRevD.71.123001}, \cite{2009Cosmic}, \cite{2012New}, \cite{2012New}, \cite{cong2014four}, \cite{Moresco_2016} and \cite{2017Age}, and are shown in Fig. \ref{fig:realohd}. Based on the real data, we can generate training data and constrain parameters with ANNs. 
	\\~	
	
\begin{figure}[t]
		\centering                                      
		\includegraphics[width=0.55\textwidth]{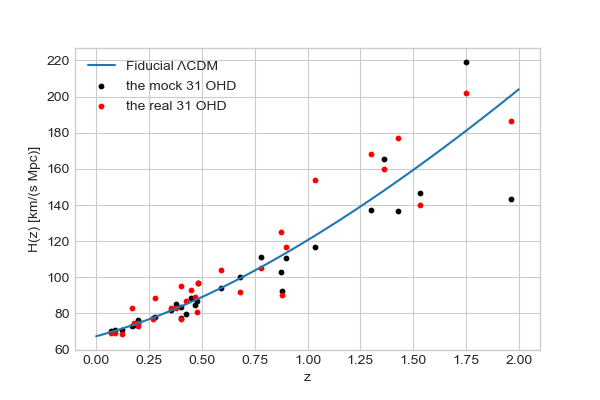}    
		\caption{31 mock datapoints are made with spatially flat $\Lambda\mathrm{CDM}$ model and the Gaussian sample. The mock OHD is based on the real OHD.}
		\label{fig:mockohd}                                     
	\end{figure}

\begin{figure}[h]
		\centering                                      
		\includegraphics[width=0.5\textwidth]{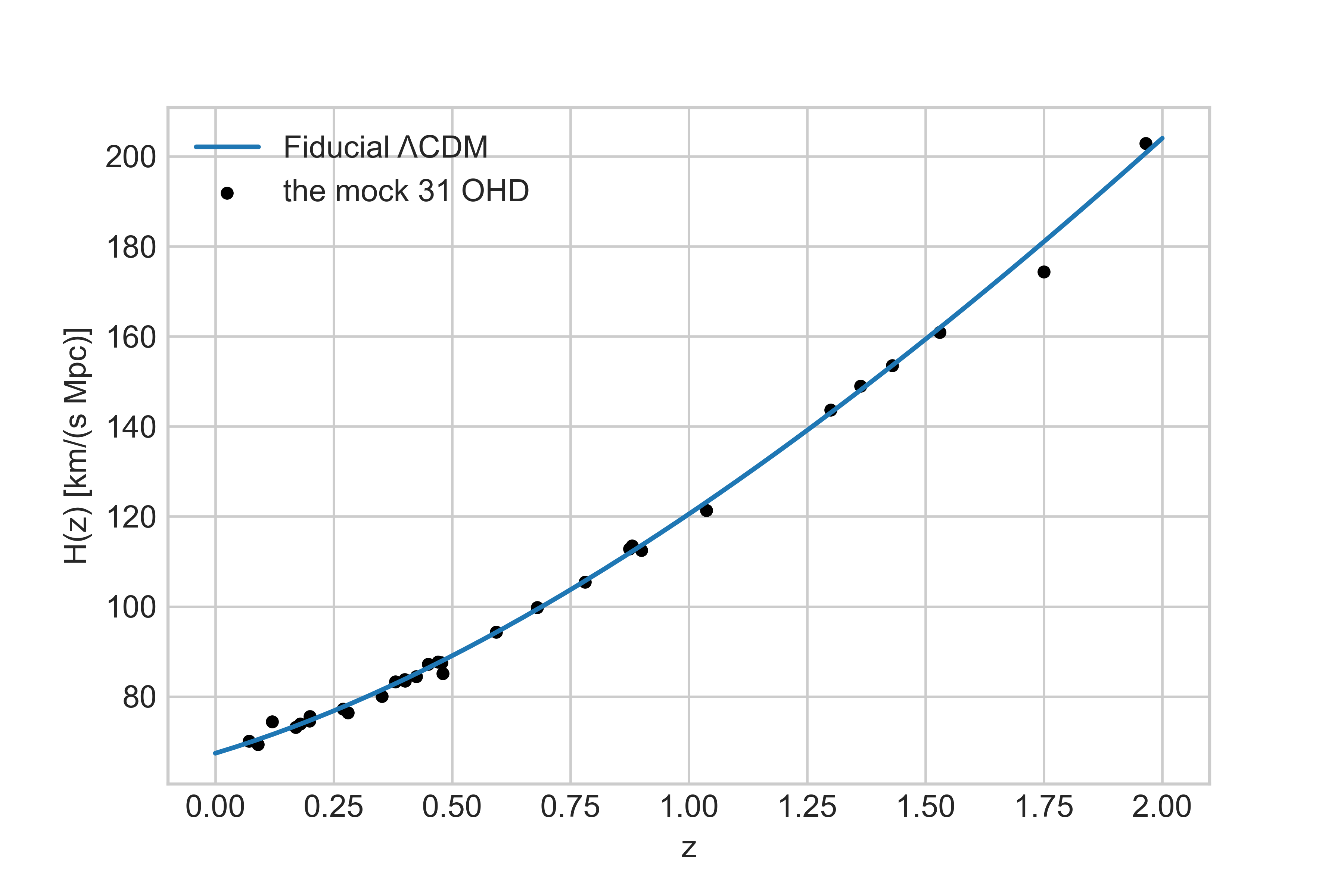}    
		\caption{The new mock OHD is also based on the real OHD. Because of the small range of the corresponding uncertainty in the Gaussian sample, the new mock 31 data points basically fit the flat $\Lambda\mathrm{CDM}$ model. }
		\label{fig:mockohd1}                                     
	\end{figure}

	According to the flat $\Lambda\mathrm{CDM}$ model, the Hubble parameter is expressed by redshift $z$ with the simple formula:
	\begin{equation}\label{eq11}
	H(z) = H_{0}\sqrt{\Omega_{m}(1+z)^{3}+\Omega_{\Lambda}} ,
	\end{equation}
	where the $H_{0}$ is the Hubble constant, or the non-flat $\Lambda\mathrm{CDM}$ model:
	\begin{equation}\label{eq11_2}
	H(z) = H_{0}\sqrt{\Omega_{m}(1+z)^{3}+\Omega_{\Lambda}+\Omega_{k}(1+z)^{2}} .
	\end{equation}
	The parameters $H_{0},\Omega_{m},\Omega_{\Lambda}$ in Eq.\ref{eq11} and Eq.\ref{eq11_2} are randomly sampled from the range [0,100], [0,1] and [0,1]. As illustrated in \cite{2021Likelihood}, when hard boundaries are added to the prior, the new posterior is almost the same as the original one, provided that the boundaries encloses the likely region of the posterior. Therefore, there is no special requirement for the sampling interval. With the random sampled parameters, as well as the $\mathbf{z} = z_i$ from the 31 OHD data, the $\textbf{\emph{H}}_{\mathrm{fid}}(z_i)$ can be easily obtained by Eq. \ref{eq11} and Eq.\ref{eq11_2}. Finally, by sampling the $\Delta \textbf{\emph{H}}_i$ in $\mathcal{N}(0,\sigma^2_i)$ \cite{2021Likelihood},\cite{yu2013nonparametric},\cite{cong2014four}, we can obtain the with the formula:
	\begin{equation}\label{eq12}
	H_{\mathrm{moc},i}=H_{\mathrm{fid}}(z_i)+\Delta H_i.
	\end{equation}
	The training data, which consists of the simulated $\textbf{\emph{H}}_{\mathrm{sim},i}$ and the corresponding $\theta=$ ($H_{0},\Omega_{m},\Omega_{\Lambda}$), should be large enough to train the ANNs, so we also set 8000 training data like \cite{2021Likelihood}. We show one of the training data in \ref{fig:mockohd}. Furthermore, in order to make a comprehensive comparison, we also simulate a new kind of mock OHD and training data by narrowing the range of the corresponding uncertainty in the Gaussian sample. One set of the mock OHD is shown in Fig. \ref{fig:mockohd1}. 
	
	\subsection{The procedure of constraining parameters} 
	The procedure of constraining $\Omega_{\Lambda},\Omega_{m},H_0$ with MAF-DAE is summarized as below: (1) Generating 8000 training data $\{\boldsymbol{\theta}, \textbf{\emph{H}}\}$ and training a DAE with the training data; (2) Generating another set of training data and encoding the $\textbf{\emph{H}}_{\mathrm{sim}}$ with the trained DAE to get lower-dimensional $\textbf{\emph{H}}_{\mathrm{sim}}$; (3) Training a MAF with the lower-dimensional $\textbf{\emph{H}}_{\mathrm{sim}}$ and corresponding parameters $\theta$; (4) Encoding the real 31 OHD with the DAE and inputting the lower-dimensional OHD to the MAF to estimate the posterior distribution $P(\theta|\textbf{\emph{H}}_{\mathrm{obs}})$.
	
	\section{MAF-IMNN for Parameter Constraint} \label{sec:s3}
	According to the method evaluating the parameters with MAF-DAE mentioned above, we apply a similar procedure to constrain the cosmological parameters with MAF-IMNN in this paper. The procedure is summarized as below: (1) Generating training data $\{\boldsymbol{\theta}, \textbf{\emph{H}}\}$ with the same model and training a IMNN; (2) Generating 8000 training data and encoding the $\textbf{\emph{H}}_{\mathrm{sim}}$ with the trained IMNN to get lower-dimensional $\textbf{\emph{H}}_{\mathrm{sim}}$; (3) Training a MAF with the lower-dimensional $\textbf{\emph{H}}_{\mathrm{sim}}$ and corresponding parameters $\theta$; (4) Encoding the real 31 OHD with the IMNN and inputting the lower-dimensional OHD to the MAF to estimate the posterior distribution $P(\theta|\textbf{\emph{H}}_{\mathrm{obs}})$. As the substitution of DAE, IMNN can find the most informative non-linear data summaries by setting fiducial parameters and calculating the Fisher information matrix on the simulated data. Althought IMNN is simulation-based, the examples proposed by \cite{2018Automatic} showed the training of the network seems fairly insensitive to the choice of fiducial parameter. In the rest of this section, we introduce the theory of IMNN briefly.
	
	\subsection{Fisher Information and compression}
	
	The Fisher information \cite{1954Statistical}, \cite{1963The}, \cite{1947Mathematics} can measure how much information that an observable variable $\mathbf{d}$ contains about parameter $\boldsymbol{\theta}$. For this reason, the larger the Fisher information is, the more informative the data is. It can be obtained by calculating the variance of the partial derivative of the natural logarithm of the likelihood $\mathcal{L}(\mathbf{d}|\boldsymbol{\theta})$ with respect to the fiducial parameter value, $\boldsymbol{\theta}^{\mathrm{fid}}$:
	\begin{equation}
	\textbf{F}_{\alpha\beta}(\boldsymbol{\theta})=-\left \langle\frac{\partial \mathrm{ln}\mathcal{L}(\mathbf{d}|\boldsymbol{\theta})}{\partial\theta_\alpha }\frac{\partial \mathrm{ln}\mathcal{L}(\mathbf{d}|\boldsymbol{\theta})}{\partial\theta_\beta }\right \rangle|_{\boldsymbol{\theta}=\boldsymbol{\theta}^{\mathrm{fid}}},
	\end{equation}
	where $\alpha, \beta \in [1,n_{\theta}]$ (where $ \alpha\neq\beta$). In our work, we used $\Lambda\mathrm{CDM}$ model, therefore $\alpha$ and $\beta$ represent $\Omega_{\Lambda},\Omega_{m}$ and $H_0$. If we use another model where theta has a higher dimension, the formula is still kept valid.
	If the likelihood is twice continuously differentiable, the expression of the Fisher information can be \cite{1963The}, \cite{1947Mathematics}, \cite{1983Theory}:
	\begin{equation}\label{eq21}
	\textbf{F}_{\alpha\beta}(\boldsymbol{\theta})=-\left \langle\frac{\partial^2 \mathrm{ln}\mathcal{L}(\mathbf{d}|\boldsymbol{\theta})}{\partial\theta_\alpha \partial\theta_\beta}\right \rangle|_{\boldsymbol{\theta}=\boldsymbol{\theta}^{\mathrm{fid}}},
	\end{equation}
	where the $\mathcal{L}(\mathbf{d}|\boldsymbol{\theta})$ is the likelihood function of the data $\mathbf{d}$ with the with $n_\mathbf{d}$ data points, and a set of $n_{\theta}$ parameters $\boldsymbol{\theta}$. We can constrain $\boldsymbol{\theta}$ in a smaller range if the $\mathcal{L}(\mathbf{d}|\boldsymbol{\theta})$ is sharp at a particular value. According to Cramér-Rao bound \citep{H1946Mathematical} \citep{calcutta1911bulletin}, under certain conditions, we can calculate the maximum Fisher information to find the minimum variance of $\theta$:
	\begin{equation}
	\left \langle(\theta_\alpha-	\left \langle \theta_{\alpha} \right \rangle)(\theta_\beta-	\left \langle\theta_\beta\right \rangle)\right \rangle \geq (\textbf{F}^{-1})_{\alpha\beta}.
	\end{equation}
	
	In particular, if the model of likelihood of the data $\mathbf{d}$ is Gaussian approximation, we can use Massively Optimised Parameter Estimation and Data (MOPED) compression algorithm \cite{10.1046/j.1365-8711.2000.03692.x} to map the data to compressed summaries. While using the MOPED, the logarithm of the likelihood should be written as 
	\begin{equation}
	-2\mathrm{ln}\mathcal{L}(\mathbf{d}|\boldsymbol{\theta})=(\mathbf{d}-\boldsymbol{\mu}(\boldsymbol{\theta}))^T \mathbf{C}^{-1}(\mathbf{d}-\boldsymbol{\mu}(\boldsymbol{\theta}))+\mathrm{ln}|2\pi \mathbf{C}|,
	\end{equation}
	where $\boldsymbol{\mu}(\boldsymbol{\theta})$ is the mean of the parameters $\boldsymbol{\theta}$ and $\mathbf{C}$ is the covariance of the data $\mathbf{d}$. Compared with MOPED, IMNN can map the data to compressed summaries without the limitation of the likelihood. $f$ is the function that transforms $n_\mathbf{d}$ data $\mathbf{d}$ to $n_\mathbf{s}$ summary $\mathbf{x}$, which means that $f : \mathbf{d}\rightarrow \mathbf{x}$. With the function $f$, the logarithm of the likelihood can be written as 
	\begin{equation}\label{eq24}
	-2\mathrm{ln}\mathcal{L}(\mathbf{x}|\boldsymbol{\theta})=(\mathbf{x}-\boldsymbol{\mu}_f(\boldsymbol{\theta}))^T \mathbf{C}_f^{-1}(\mathbf{x}-\boldsymbol{\mu}_f(\boldsymbol{\theta})),
	\end{equation}
	where 
	\begin{equation}\label{eq25}
	\boldsymbol{\mu}_f(\boldsymbol{\theta})=\frac{1}{n_s}\sum_{i=1}^{n_s}x^{s}_i,
	\end{equation}
	is the mean value of $n_\mathbf{s}$ summaries $\{x^s_i|i\in[1,n_\mathbf{s}]\}$, and $\mathbf{C}^{-1}_f$ is the inverse of the covariance matrix:
	\begin{equation}\label{eq26}
	(\mathbf{C}_f)_{\alpha\beta}=\frac{1}{n_s-1}\sum_{i=1}^{n_s}(\mathbf{x}^s_i-\boldsymbol{\mu}_f)_{\alpha}(\mathbf{x}^s_i-\boldsymbol{\mu}_f)_{\beta}.
	\end{equation}
	
	\begin{figure}
		\centering                                      
		\includegraphics[width=0.55\textwidth]{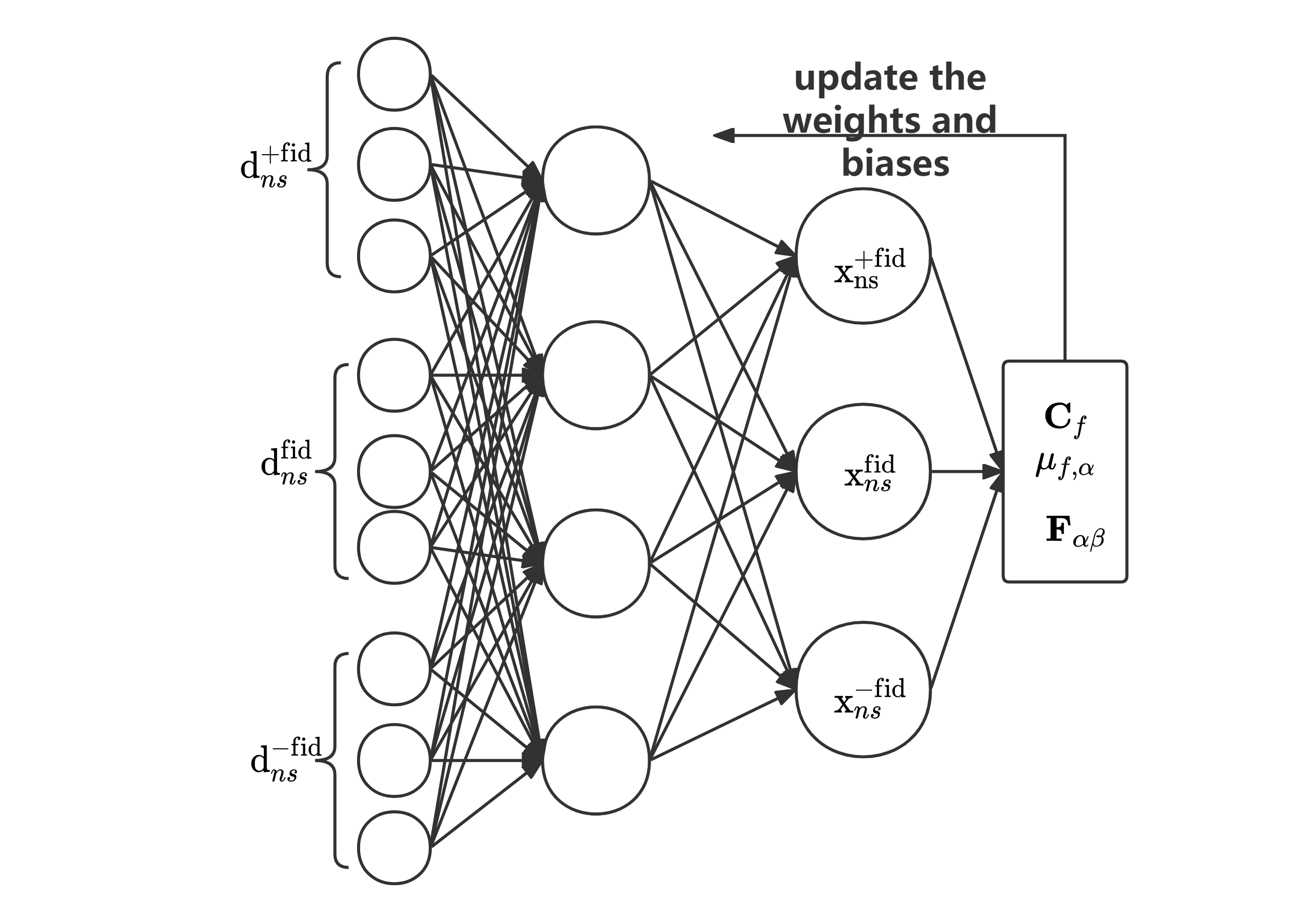}    
		\caption{The concise structure of the IMNN.  The ANN can compress the input data $\mathbf{d}$ to the $\mathbf{x}$. The loss function is calculated with $\mathbf{C}_f,\mu_{f,\alpha}$ and $\textbf{F}_{\alpha\beta}$ with $\mathbf{x}^{+\mathrm{fid}}$. Normally $\mathbf{x}$ would be considered as the network output, but we can also choose Fisher information matrix as the network output, which means that $\mathbf{x}$ will be the intermediate output of the neural network before calculating the loss function. In our work, obviously we wanted to obtain $\mathbf{x}$.}
		\label{fig:imnn}                                     
	\end{figure}
	
	While training, each summary $\mathbf{x}^s_i$ is obtained from $f:\mathbf{d}^s_i\rightarrow \mathbf{x}^s_i$, where $\boldsymbol{d}^s_i$ is from the simulation $\mathbf{d}^s_i=\mathbf{d}^s(\boldsymbol{\theta},i)$ at the fiducial values $\boldsymbol{\theta}$. With Eq. (\ref{eq21}) and Eq. (\ref{eq24}), the Fisher information matrix can be expressed in the form:
	\begin{equation}\label{eq27}
	\textbf{F}_{\alpha\beta}=Tr[\boldsymbol{\mu}_{f,a}^{\,\,\,\,T} \mathbf{C}^{-1}_f \boldsymbol{\mu}_{f,\beta}].
	\end{equation}
	The $\boldsymbol{\mu}_{f,a}$ can be calculated by 
	\begin{equation}
	\boldsymbol{\mu}_{f,\alpha}=\frac{\partial}{\partial\theta_{\alpha}}\frac{1}{n_s}\sum_{i=1}^{n_s}x^{s\,\mathrm{fid}}_i=\frac{1}{n_s}\sum_{i=1}^{n_s}\left(\frac{\partial x}{\partial\theta_\alpha}\right)^{s\,\mathrm{fid}}_i.
	\end{equation}
	Note that the fiducial parameters are only used in the simulations, so we need to do some additional numerical differentiation to calculate $\left(\frac{\partial x}{\partial\theta_\alpha}\right)^{s\,\mathrm{fid}}_i$ with these three copies of the simulation, $\mathbf{d}^{s\, \mathrm{fid}}_i=\mathbf{d}^s(\boldsymbol{\theta}^{\mathrm{fid}},i)$, $\mathbf{d}^{s\,fid-}_i=\mathbf{d}^s(\boldsymbol{\theta}^{\mathrm{fid}}-\Delta\boldsymbol{\theta}^-,i)$ and $\mathbf{d}^{s\,fid+}_i=\mathbf{d}^s(\boldsymbol{\theta}^{\mathrm{fid}}+\Delta\boldsymbol{\theta}^+,i)$, where the $\Delta\boldsymbol{\theta}^{ \pm}$ is the small deviation from the fiducial parameter values. With the above conditions, the $\left(\frac{\partial x}{\partial\theta_\alpha}\right)^{s\,\mathrm{fid}}_i$ is therefore given by
	\begin{equation}
	\left(\frac{\partial x}{\partial\theta_\alpha}\right)^{s\,\mathrm{fid}}_i\approx\frac{x^{s\,fid+}_i-x^{s\,fid-}_i}{\Delta\theta^+_\alpha-\Delta\theta^-_\alpha}.
	\end{equation}
	
	Also we can calculate the $\left(\frac{\partial x}{\partial\theta_\alpha}\right)^{s\,\mathrm{fid}}_i$ with the formula
	\begin{equation}
	\boldsymbol{\mu}_{f,\alpha}=	\frac{1}{n_s}\sum_{i=1}^{n_s}\sum_{k=1}^{n_d}\frac{\partial x^{s\,\mathrm{fid}}_{ik}}{\partial d_k}\frac{\partial d^{s\,\mathrm{fid}}_{ik}}{\partial\theta_\alpha},
	\end{equation}
	where $i$ represents the random initialisation of the simulation, and $k$ represents the data point in the simulation.
	
	Here, both the values of $\boldsymbol{\mu}_{f,a}$ and $\mathbf{C}^{-1}_f$ are calculated with fixed, fiducial parameter values, $\boldsymbol{\theta}^{\mathrm{fid}}$. In IMNN, the function $f$ is a neural network, which will be described in the next subsection.
	
	\subsection{Implementing $f$ with artificial neural networks}
	
	A basic neuron unit is in the form:
	\begin{equation}\label{eq31}
	a^l_j=\phi(\sum_{j}w^l_{ji}a^{l-1}_i+b^l_j).
	\end{equation}
	The loss function in IMNN is defined using the Fisher information matrix $|\mathbf{F}|$:
	\begin{equation}
	\Lambda=-\frac{1}{2}|\mathbf{F}|^2
	\end{equation}
	or
	\begin{equation}
	\frac{\partial\Lambda}{\partial \mathbf{a}^L}=-|\mathbf{F}|+|\mathbf{C}_f|.
	\end{equation}
	With the loss function, the weights and biases will be updated by gradient descent  \citep{2001Convergence} in the updating procedure:
	\begin{equation}
	w^l_{ji}\rightarrow w^l_{ji}-\eta \frac{\partial \Lambda}{\partial w^l_{ji}}
	\end{equation}
	and
	\begin{equation}
	b^l_i\rightarrow b^l_i-\eta \frac{\partial \Lambda}{\partial b^l_i},
	\end{equation}
	where $\eta$ is the learning rate, which controls the size of the steps in the procedure of updating the weights and biases \citep{theodoridis2015neural}. The $i$ means the $i^{th}$ element of the output vector of a collections of neurons in the $(l-1)^{th}$ layer, while the $j$ means the $j^{th}$ neuron in the $l^{th}$ layer. The mean $\boldsymbol{\mu}_f$, covariance $\mathbf{C}^f$, which can be calculated with the Eq. (\ref{eq25}) and Eq. (\ref{eq26}), are part of the loss function and therefore are functions of the weights and biases. The concise structure of the IMNN is shown in Fig. \ref{fig:imnn}. 
	
	\section{CONSTRAINTS WITH REAL OHD} \label{sec:s4}
\begin{figure*}
	\centering                                      
	\includegraphics[width=1.1\textwidth]{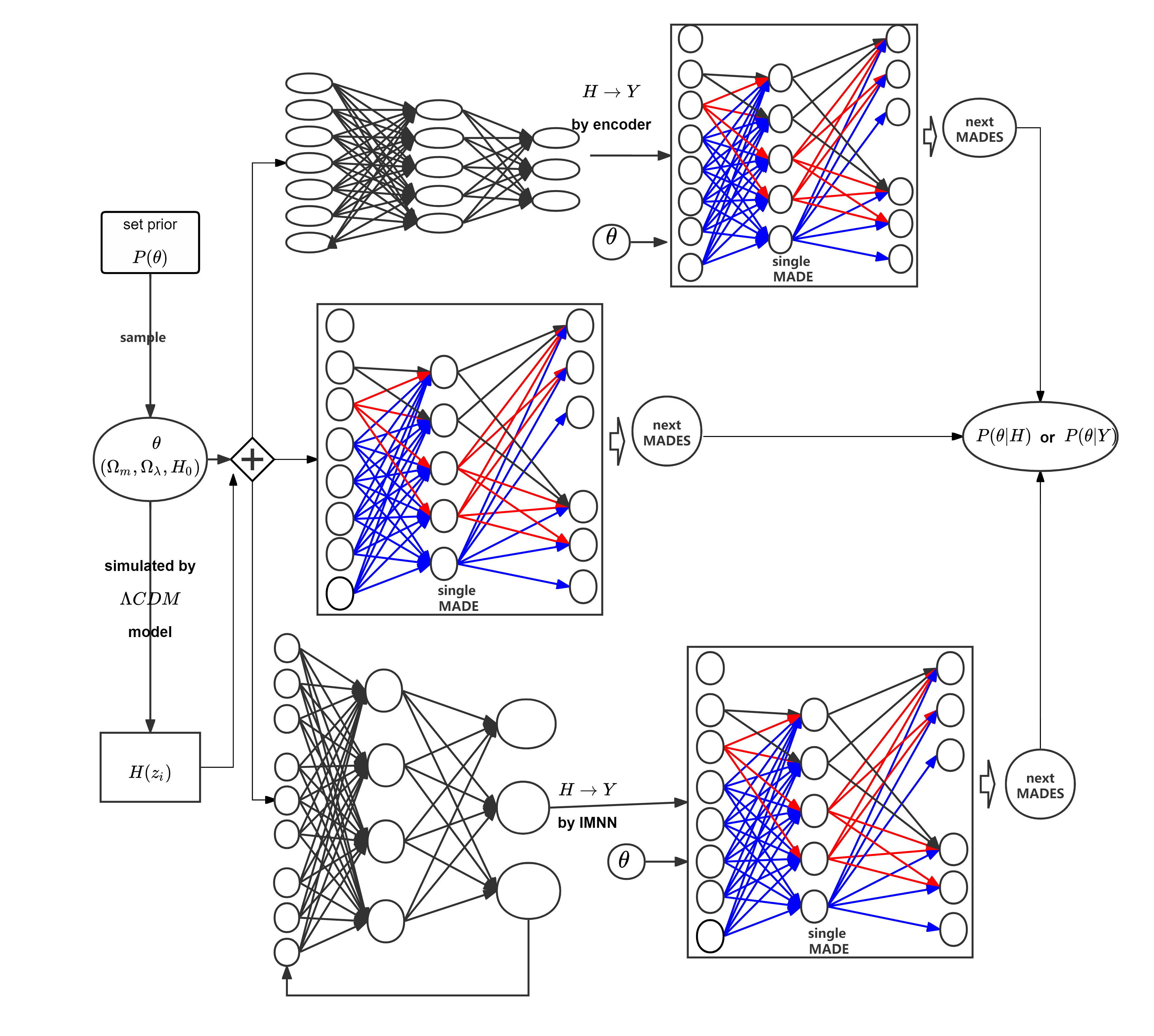}    
	\caption{The procedure in this work. We first produce training data. The upper part is MAF-DAE. The  training data is compressed into low-dimensional $\textbf{\emph{y}}$ by IMNN, then $\textbf{\emph{y}}$ and the corresponding $\boldsymbol{\theta}$ are transmitted to the MAF. The middle part is  MAF, which is trained with original-dimensional training data. The lower part is MAF-IMNN. In this method, the training data is compressed into low-dimensional $\textbf{\emph{y}}$ by DAE. Then $\textbf{\emph{y}}$ and the corresponding $\boldsymbol{\theta}$ are transmitted to the MAF. After training, when evaluating cosmological parameters, MAFs can give posterior $P(\boldsymbol{\theta}|\textbf{\emph{H}}_{obs})$ and $P(\boldsymbol{\theta}|\boldsymbol{y}_{obs})$ (or $P(\boldsymbol{\theta}|\textbf{\emph{H}}_{moc})$ and $P(\boldsymbol{\theta}|\boldsymbol{y}_{moc})$).}
	\label{fig:procedure}                                     
	\end{figure*}

	In this work, we use 3 types of methods (shown in Fig. \ref{fig:procedure}) for $\Lambda\mathrm{CDM}$s with and without curvature to constrain the cosmological parameters, which are: (1) using MAF and $\textbf{\emph{H}}_{\mathrm{obs}}$ to estimate the posterior distribution $P(\boldsymbol{\theta}|\textbf{\emph{H}}_{\mathrm{obs}})$ directly. (2) using the MAF-DAE to estimate the posterior distribution $P(\boldsymbol{\theta}|\textbf{\emph{H}}_{\mathrm{obs}})$ with $\textbf{\emph{H}}_{\mathrm{obs}}$. (3) using MAF-IMNN to estimate the posterior distribution $P(\boldsymbol{\theta}|\textbf{\emph{H}}_{\mathrm{obs}})$ with $\textbf{\emph{H}}_{\mathrm{obs}}$. We used the results from MAF as reference.
	
	\begin{figure}[h]
		\centering                                      
		\includegraphics[width=0.45\textwidth]{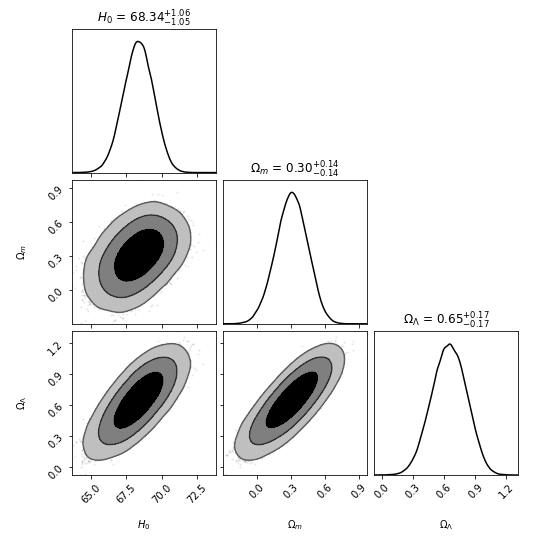}    
		\caption{The posterior distribution estimated by MAF, the real 31 OHD and non-flat $\Lambda\mathrm{CDM}$ model.}
		\label{fig:maf k+m+l}                                     
	\end{figure}
	
	\begin{figure}[h]
		\centering                                      
		\includegraphics[width=0.45\textwidth]{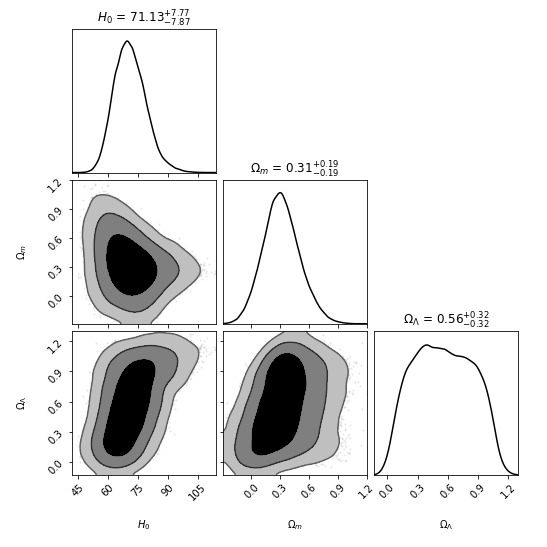}    
		\caption{The posterior distribution estimated by MAF-IMNN, the real OHD and non-flat $\Lambda\mathrm{CDM}$ model. The OHD is compressed into 10 dimension.}
		\label{fig:10imnn k+m+l}                                     
	\end{figure}
	
	\begin{figure}[h]
		\centering                                      
		\includegraphics[width=0.45\textwidth]{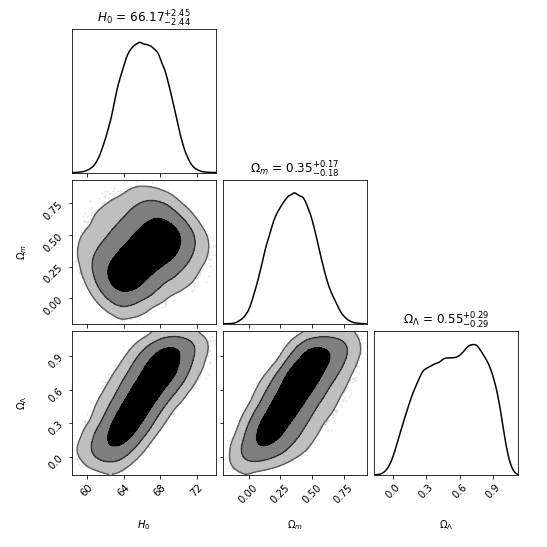}    
		\caption{The posterior distribution estimated by MAF-DAE, the real OHD and non-flat $\Lambda\mathrm{CDM}$ model. The OHD is compressed into 10 dimension.}
		\label{fig:10ae k+m+l}                                     
	\end{figure}
	
	\begin{figure}[h]
		\centering                                      
		\includegraphics[width=0.45\textwidth]{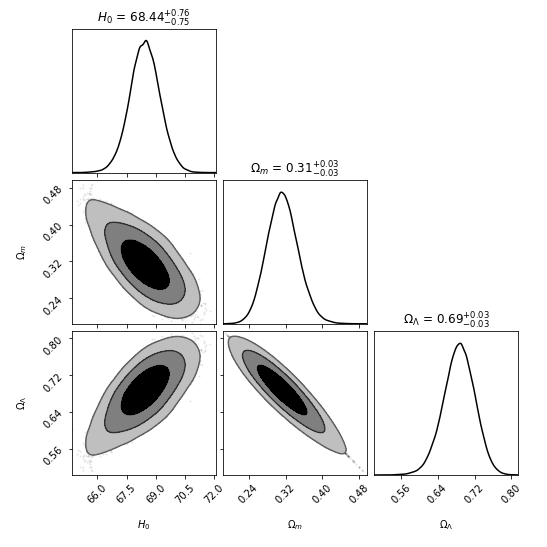}    
		\caption{The posterior distribution estimated by MAF, the real 31 OHD and flat $\Lambda\mathrm{CDM}$ model.}
		\label{fig:maf m+l}                                     
	\end{figure}
	
	\begin{figure}[h]
		\centering                                      
		\includegraphics[width=0.45\textwidth]{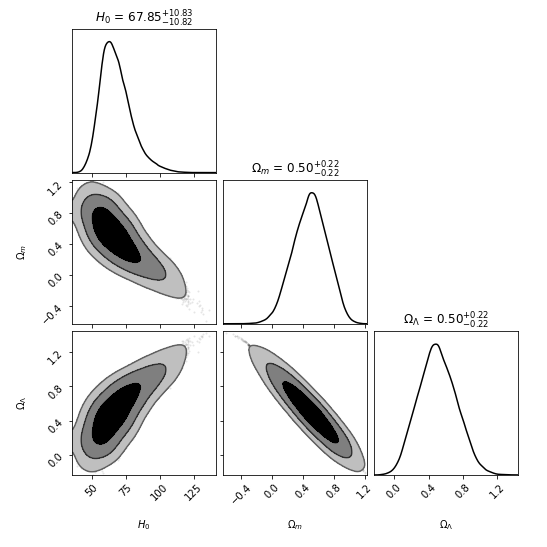}    
		\caption{The posterior distribution estimated by MAF-IMNN, the real OHD and flat $\Lambda\mathrm{CDM}$ model. The OHD is compressed into 10 dimension.}
		\label{fig:10imnn m+l}                                     
	\end{figure}
	
	\begin{figure}[h]
		\centering                                      
		\includegraphics[width=0.45\textwidth]{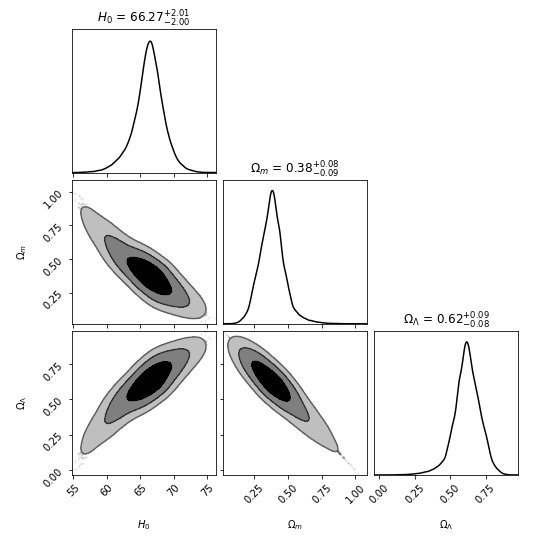}    
		\caption{The posterior distribution estimated by MAF-DAE, the real OHD and flat $\Lambda\mathrm{CDM}$ model. The OHD is compressed into 10 dimension.}
		\label{fig:10ae m+l}                                     
	\end{figure}
	
	With $\textbf{\emph{H}}_{\mathrm{obs}}$ and the non-flat $\Lambda\mathrm{CDM}$ model, the posterior distribution estimated by MAF gives $H_0=68.34^{+1.06}_{-1.05}$ km s$^{-1}$ Mpc$^{-1}$, $\Omega_{m}=0.30^{+0.14}_{-0.14}$, $\Omega_{\Lambda}=0.65^{+0.17}_{-0.17}$, the posterior distribution estimated by MAF-IMNN gives $H_0=71.13^{+7.77}_{-7.87}$ km s$^{-1}$ Mpc$^{-1}$, $\Omega_{m}=0.31^{+0.19}_{-0.19}$, $\Omega_{\Lambda}=0.65^{+0.17}_{-0.17}$, the posterior distribution estimated by MAF-DAE gives $H_0=66.17^{+2.45}_{-2.44}$ km s$^{-1}$ Mpc$^{-1}$, $\Omega_{m}=0.35^{+0.17}_{-0.18}$, $\Omega_{\Lambda}=0.55^{+0.29}_{-0.29}$. 
	
	Meanwhile, with $\textbf{\emph{H}}_{\mathrm{obs}}$ and the flat $\Lambda\mathrm{CDM}$ model, the posterior distribution estimated by MAF gives $H_0=68.44^{+0.76}_{-0.75}$ km s$^{-1}$ Mpc$^{-1}$, $\Omega_{m}=0.31^{+0.03}_{-0.03}$, $\Omega_{\Lambda}=0.69^{+0.03}_{-0.03}$, the posterior distribution estimated by MAF-IMNN gives $H_0=67.85^{+10.83}_{-10.82}$ km s$^{-1}$ Mpc$^{-1}$, $\Omega_{m}=0.50^{+0.22}_{-0.22}$, $\Omega_{\Lambda}=0.50^{+0.22}_{-0.22}$, the posterior distribution estimated by MAF-DAE gives $H_0=66.27^{+2.01}_{-2.00}$ km s$^{-1}$ Mpc$^{-1}$, $\Omega_{m}=0.38^{+0.08}_{-0.09}$, $\Omega_{\Lambda}=0.62^{+0.09}_{-0.08}$. We showed the table and figures of these posterior distributions in Fig. \ref{fig:maf k+m+l}, \ref{fig:10imnn k+m+l}, \ref{fig:10ae k+m+l}, \ref{fig:maf m+l}, \ref{fig:10imnn m+l}, \ref{fig:10ae m+l} and Table \ref{tab:posterior distribution}. 
	
	\begin{table}[h!]
		\caption{The posterior distribution}
		\begin{tabular}{|l|c|r|c|} 
			\hline
			 & \emph{$H_0$} & \emph{$\Omega_{m}$} & \textbf{$\Omega_{\lambda}$} \\
			\hline
						&&&\\
			non-flat $\Lambda\mathrm{CDM}$& &  &  \\
			\hline
			&&&\\
			MAF&$68.34^{+1.06}_{-1.05}$& $0.30^{+0.14}_{-0.14}$& $0.65^{+0.17}_{-0.17}$\\
			&&&\\
			\hline
			&&&\\
			MAF-IMNN&$71.13^{+7.77}_{-7.87}$&$0.31^{+0.19}_{-0.19}$&$0.65^{+0.17}_{-0.17}$\\
			&&&\\
			\hline
			&&&\\
			MAF-DAE&$66.17^{+2.45}_{-2.44}$&$0.35^{+0.17}_{-0.18}$&$0.55^{+0.29}_{-0.29}$ \\
			&&&\\
			\hline
						&&&\\
			flat $\Lambda\mathrm{CDM}$& &  & \\
			\hline
			&&&\\
			MAF&$68.44^{+0.76}_{-0.75}$ &$0.31^{+0.03}_{-0.03}$ &$0.69^{+0.03}_{-0.03}$\\
			&&&\\
			\hline
			&&&\\
			MAF-IMNN&$67.85^{+10.83}_{-10.82}$ &$0.50^{+0.22}_{-0.22}$& $0.50^{+0.22}_{-0.22}$\\
			&&&\\
			\hline
			&&&\\
			MAF-DAE&$66.27^{+2.01}_{-2.00}$& $0.38^{+0.08}_{-0.09}$ &$0.62^{+0.09}_{-0.08}$\\
			&&&\\
			\hline
		\end{tabular}
		\label{tab:posterior distribution}
	\end{table}
	
	\section{THE COMPARISON OF IMNN AND DAE} \label{sec:s5}
	
	To avoid computationally expensive calculation in comparing posterior directly, we apply some criteria, which can be calculated by posterior distributions. We try to train both DAE and IMNN to compress the 31 dimensional $\textbf{\emph{H}}_{\mathrm{obs}}$ into different dimensions and estimate the posterior $P(\boldsymbol{\theta}|\textbf{\emph{H}}_{\mathrm{obs}})$ in different learning rates, so that we can compare the results under different learning rates and dimensionality reduction processes. In addition, we take the posterior $P_1(\boldsymbol{\theta}|\textbf{\emph{H}}_{\mathrm{obs}})$ obtained from only MAF as the standard posterior in an effort to investigate the impact of the addition of DAE and IMNN on the standard posterior. In the following subsection, we introduce the criteria we apply in this work and do the comparison of DAE and IMNN.  
	
	\subsection{Comparison criteria}
	
	In this paper, we apply two criteria, KL divergence and figure of merit(FoM).
	
	\emph{Kullback-Leibler divergence (KL divergence)}. Kullback–Leibler divergence is a statistical distance which can measure how one probability distribution is different from a second one. That is, Kullback–Leibler divergence can be used to calculate how much information is lost when we approximate one distribution with another. Generally, while processing probability and statistics, we can replace the observed data or complex distribution with a simpler approximate distribution.  Suppose that there are two probability density distributions red $p_1(\boldsymbol{\theta})$ and $p_2(\boldsymbol{\theta})$, where  $p_2(\boldsymbol{\theta})$ is the simulation of the $p_1(\boldsymbol{\theta})$. Then we can use the KL divergence to calculate the information loss of approximating $p_1(\boldsymbol{\theta})$ using $p_2(\boldsymbol{\theta})$. In this case, The KL divergence from $p_1(\boldsymbol{\theta})$ to $p_2(\boldsymbol{\theta})$ is defined as
	\begin{equation}
	D_{\mathrm{KL}}=(p_1(\boldsymbol{\theta})||p_2(\boldsymbol{\theta}))=\mathbb{E}_{p_1(\boldsymbol{\theta})}(\log\,p_1(\boldsymbol{\theta})-\log\,p_2(\boldsymbol{\theta})).
	\end{equation}
	In this paper, we sample $M$ samples $\{\boldsymbol{\theta}_i \}$ from the posterior, so the KL divergence is estimated with:
	\begin{equation}\label{eqkl}
	D_{\mathrm{KL}}(p_1||p_2)=\frac{1}{M}\sum_{i=1}^{M}(\mathrm{ln}\,P_1(\boldsymbol{\theta}_i|\textbf{\emph{H}}_{\mathrm{obs}})-\mathrm{ln}\,P_2(\boldsymbol{\theta}_i|\textbf{\emph{H}}_{\mathrm{obs}})),
	\end{equation} 
	where $P_2(\boldsymbol{\theta}|\textbf{\emph{H}}_{\mathrm{obs}})$ is the posterior calculated from MAF-DAE or MAF-IMNN and $P_1(\boldsymbol{\theta}|\textbf{\emph{H}}_{\mathrm{obs}})$ is the posterior calculated with only MAF. From Eq. (\ref{eqkl}), it is obvious that the smaller the KL divergence, the closer the $P_1(\boldsymbol{\theta}|\textbf{\emph{H}}_{\mathrm{moc}})$ and $P_2(\boldsymbol{\theta}|\textbf{\emph{H}}_{\mathrm{moc}})$. When $D_{\mathrm{KL}}(p_1||p_2)=0$, it means that the two posterior are almost identical.
	
	\emph{Figure of merit}(FoM). When constraining the parameters, we want to get an accurate range of the parameters and tighten the constraints. The FoM used in this work is similar to the one adopted by \cite{Ma_2011} and \cite{2011Constraints} in their work. The FoM is defined as:
	\begin{equation}
	P(\boldsymbol{\theta}|\textbf{\emph{H}}_{\mathrm{obs}})=\mathrm{const.}=\exp(-\Delta \mathcal{X}^2/2)P_{\mathrm{max}},
	\end{equation}
	where $P_{\mathrm{max}}$ is the maximum probability density of the posterior, and $\exp(-\Delta \mathcal{X}^2/2)$ is a constant which ensures that $\exp(-\Delta \mathcal{X}^2/2)P_{\mathrm{max}}$ is equal to the probability density at the boundary of the 95.44\% confidence region of the Gaussian distribution. According to \cite{2021Likelihood}, $\exp(-\Delta \mathcal{X}^2/2)$ here takes the same value of 8.02. The FoM represents the reciprocal volume of the confidence region of the posterior, so the larger the FoM, the tighter the constraint of the parameters are.
	
	\subsection{Experiments and results}
	
	\subsubsection{Comparison using KL divergence}
	
\begin{figure}
		\centering                                      
		\includegraphics[width=0.5\textwidth]{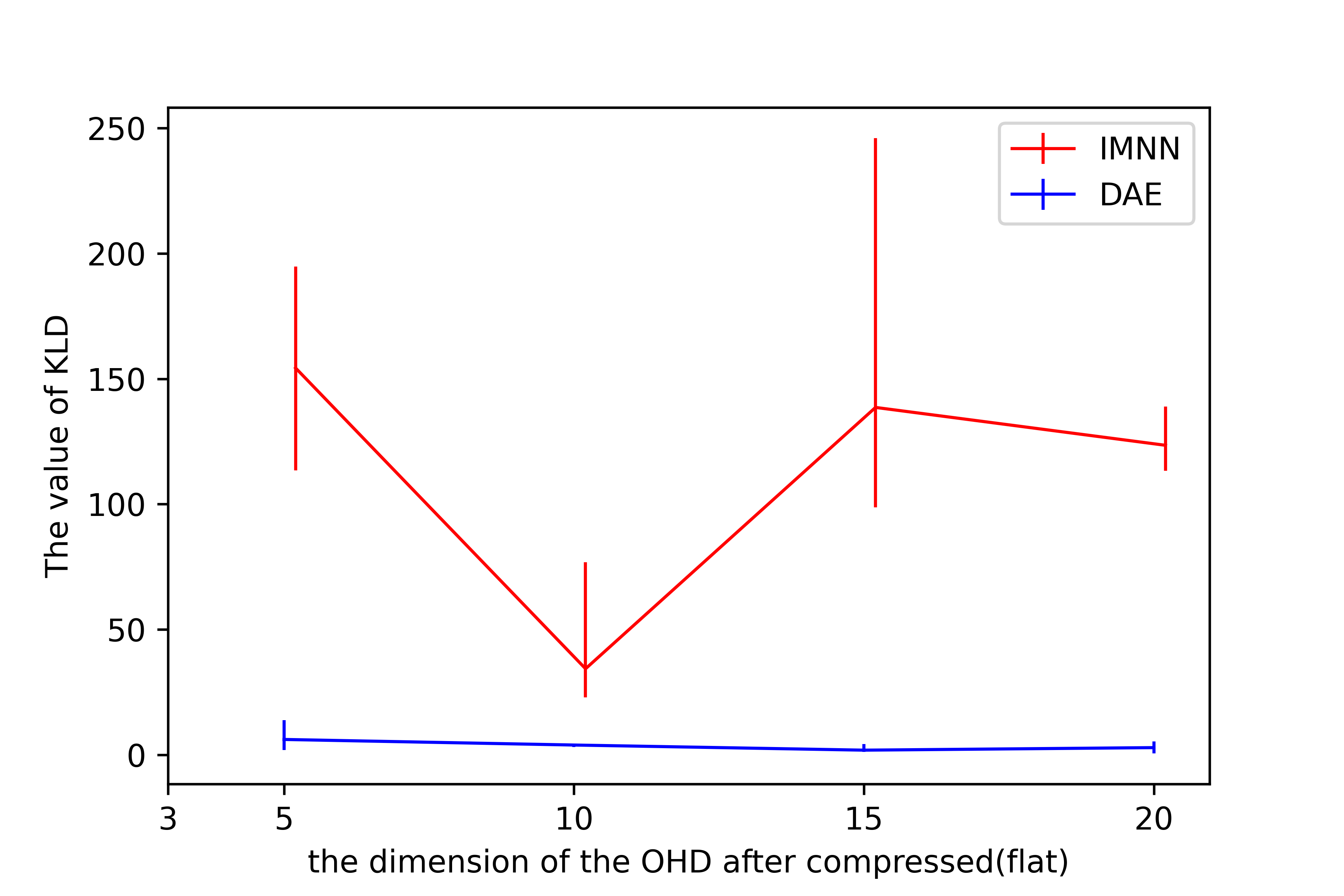}    
		\caption{KL divergence calculated by MAF-IMNN or MAF-DAE in the flat $\Lambda\mathrm{CDM}$ model. }
		\label{fig:ml-kld}                                     
	\end{figure}

\begin{figure}
		\centering                                      
		\includegraphics[width=0.5\textwidth]{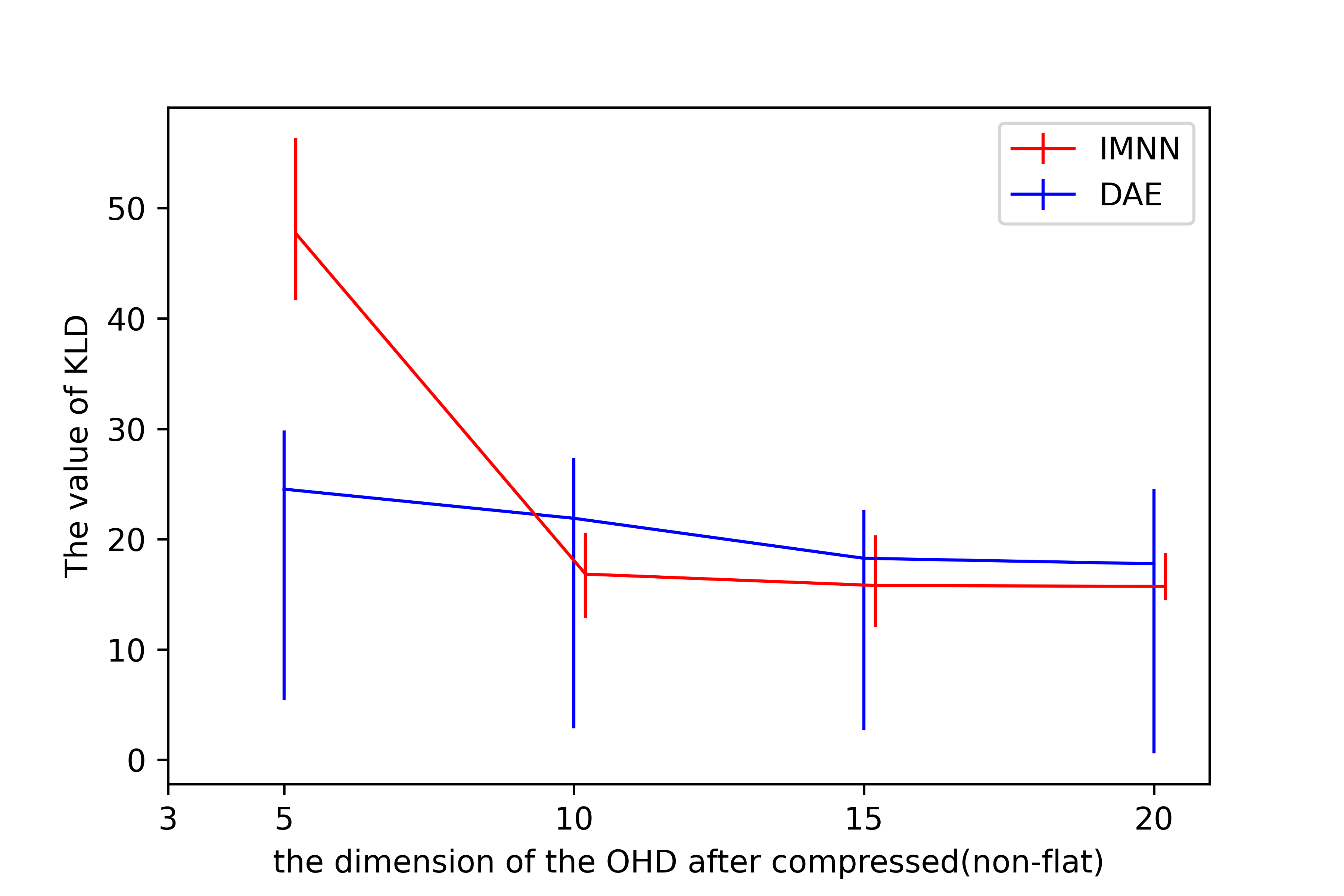}    
		\caption{KL divergence calculated by MAF-IMNN or MAF-DAE in the non-flat $\Lambda\mathrm{CDM}$ model. }
		\label{fig:kml-kld}                                     
	\end{figure}

\begin{figure}
		\centering                                      
		\includegraphics[width=0.5\textwidth]{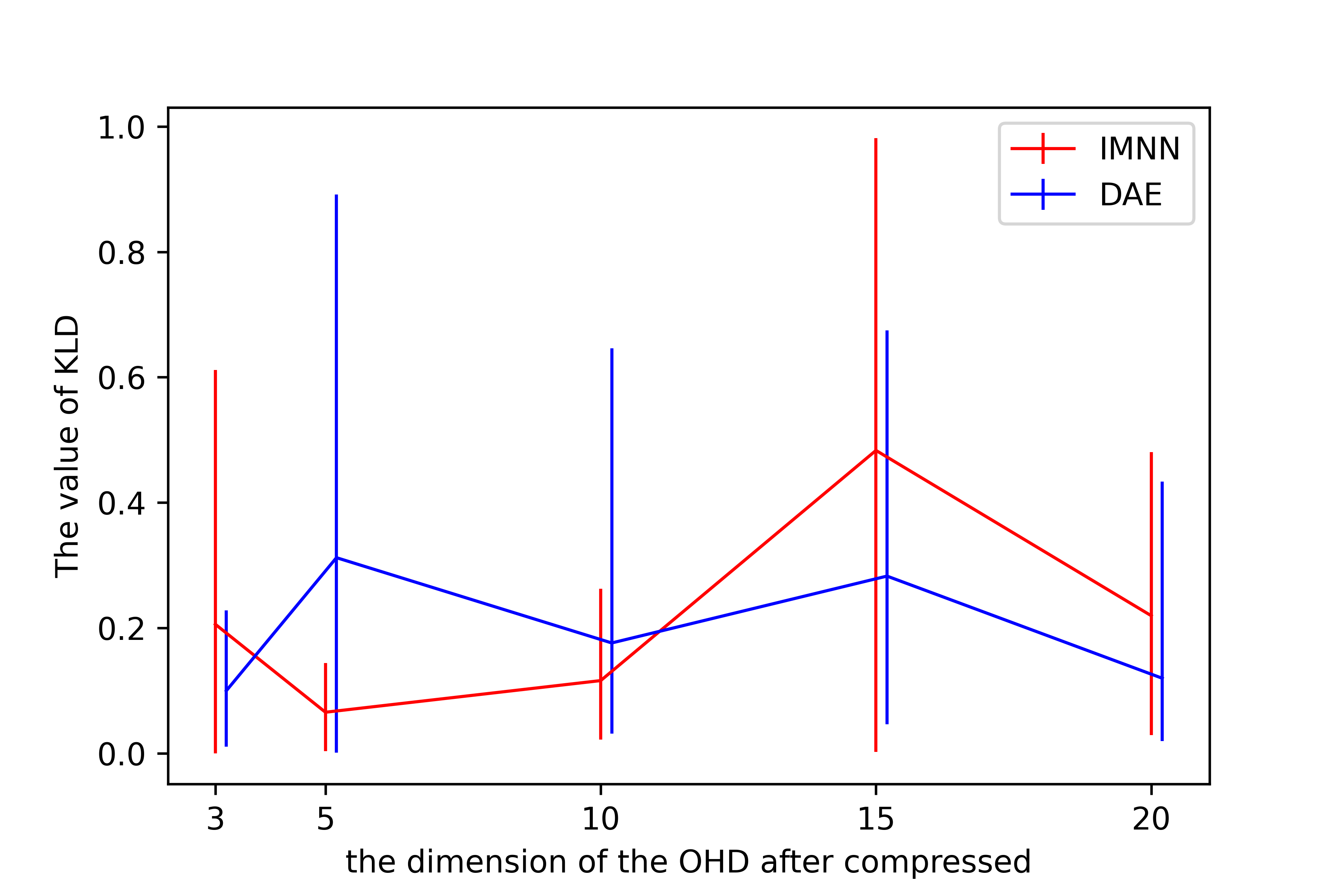}    
		\caption{KL divergence calculated by MAF-IMNN or MAF-DAE with low uncertainty. }
		\label{fig:smallkld}                                     
	\end{figure}

	We show the different Fom in Fig. \ref{fig:ml-kld}, Fig. \ref{fig:kml-kld} and Fig. \ref{fig:smallkld}. The results show signs that DAE could make a better performance than IMNN. Besides, MAF-IMNN and MAF-DAE have better results in the non-flat $\Lambda\mathrm{CDM}$, as is showed in Fig. \ref{fig:kml-kld}, the KL divergence increases with the dimension reduction.
	
	\subsubsection{Comparison using FoM}
	
\begin{figure}
		\centering                                      
		\includegraphics[width=0.5\textwidth]{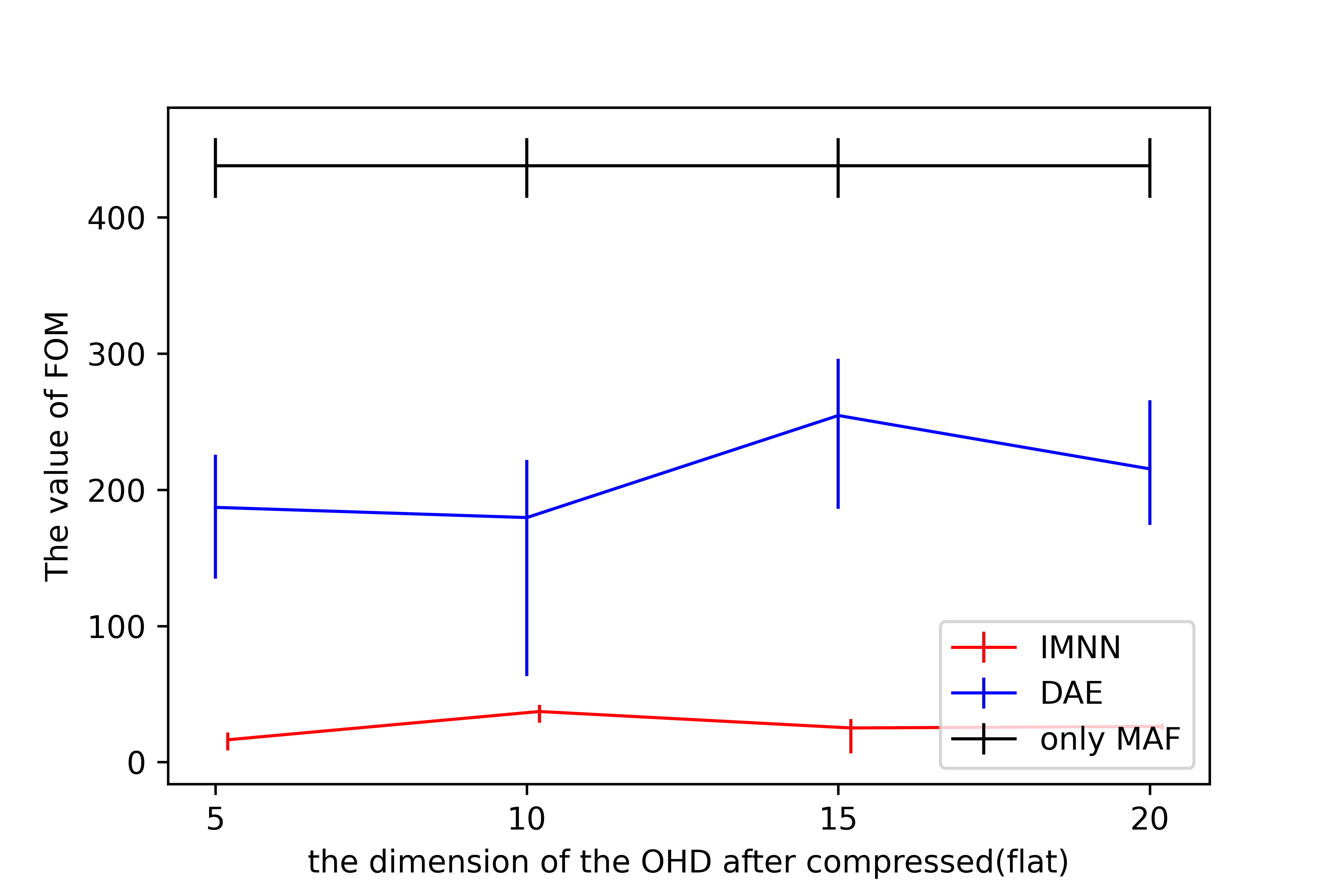}    
		\caption{FoM calculated by MAF-IMNN and MAF-DAE in the flat $\Lambda\mathrm{CDM}$ model. The black line is the FoM of the posterior calculated by only MAF with the uncompressed mock OHD in different learning rates.}
		\label{fig:mlfom}                                     
	\end{figure}

\begin{figure}
		\centering                                      
		\includegraphics[width=0.5\textwidth]{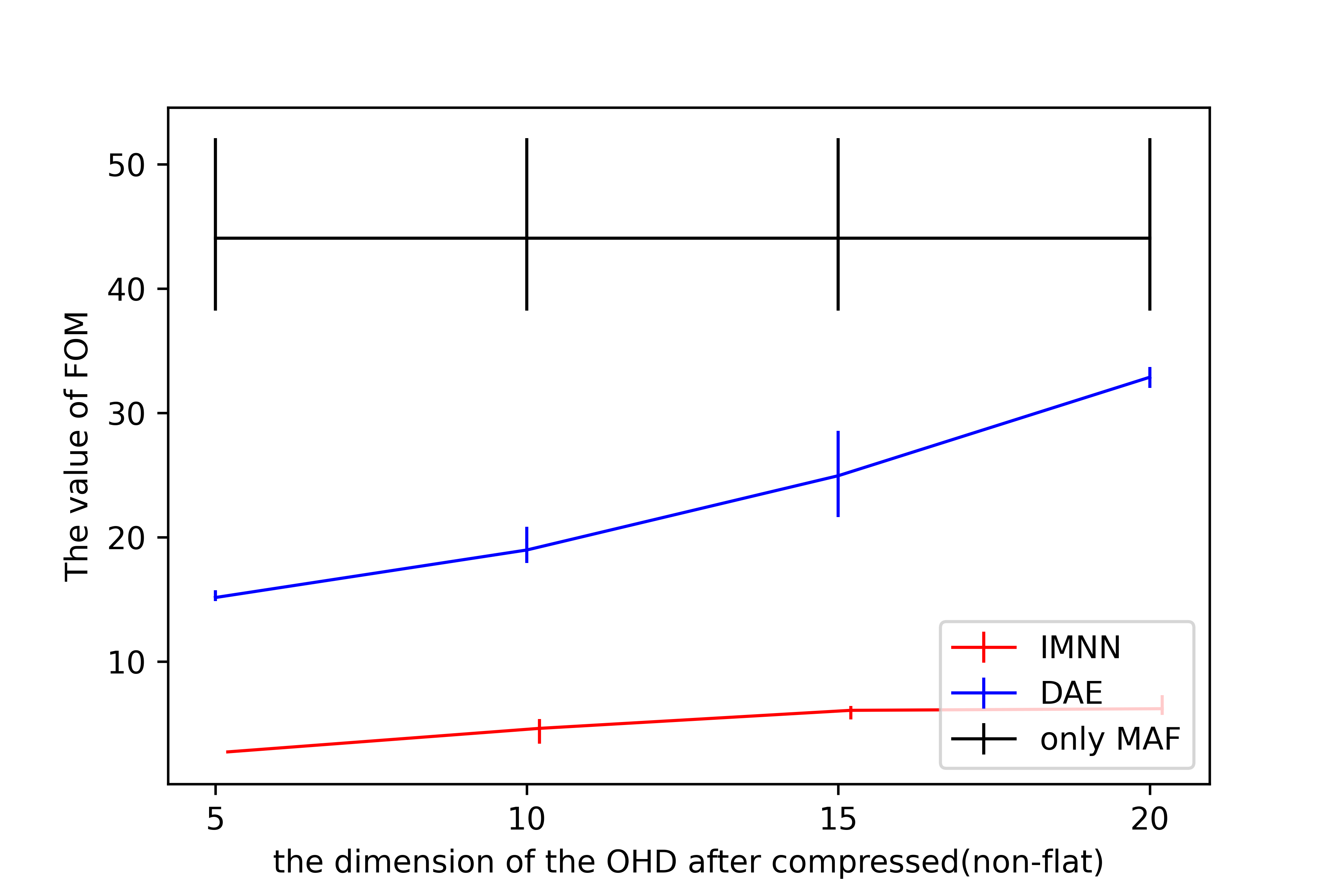}    
		\caption{FoM calculated by MAF-IMNN and MAF-DAE in the non-flat $\Lambda\mathrm{CDM}$ model. The black line is the FoM of the posterior calculated by only MAF with the uncompressed mock OHD in different learning rates.}
		\label{fig:kmlfom}                                     
	\end{figure}

\begin{figure}
		\centering                                      
		\includegraphics[width=0.5\textwidth]{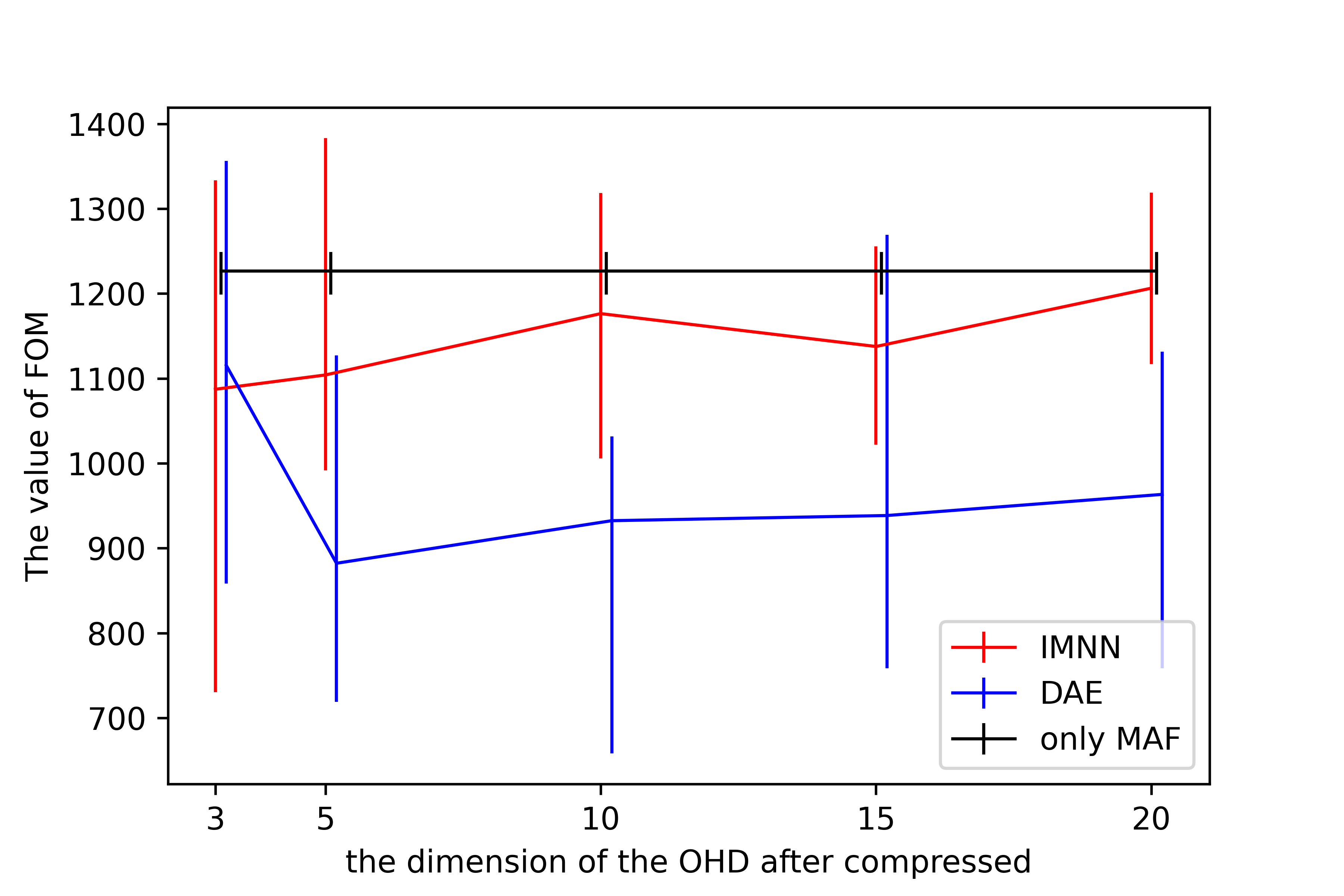}    
		\caption{FoM calculated by MAF-IMNN and MAF-DAE in the low uncertainty. The black line is the FoM of the posterior calculated by only MAF with the uncompressed mock OHD in different learning rates.}
		\label{fig:smallfom}                                     
	\end{figure}

	We show the different Fom in Fig. \ref{fig:mlfom}, Fig. \ref{fig:kmlfom} and Fig. \ref{fig:smallfom}. We can see that the FoM calculated from the posterior from MAF-DAE is generally larger, meaning that the data processed by DAE can give a tighter posterior. While using the small error training data, MAF-IMNN gives a bit smaller distributions.
	
	\section{CONCLUSIONS AND DISCUSSION}\label{sec:s6}
	
	In this paper, we validate the feasibility of MAF-IMNN, and compare IMNN and DAE in the procedure of constraining the cosmological parameters. Since we have already demonstrated that the confidence regions estimated with MAF are very close to those of MCMC \cite{2021Likelihood}, and the purpose of this work is to compare IMNN and DAE, we therefore used the results of MAF as the standard and did not calculate the KL divergence and FoM of the MCMC results.
	
	We also used different model to simulate the training data to do a comprehensive comparison between DAE and IMNN. With the small error training data, The performance of those two methods is very similar. With the normal training data, the overall performance of DAE is better than that of IMNN. Nevertheless, there is always an apparent influence from IMNN or DAE, no matter which kind of training data was uesd. We can also estimate another cosmological model as long as we generate the training data according to the cosmological model.
	
	Admittedly, our work is not perfect in some aspects. Firstly, the simulation model in this paper is not complex enough to simulate the generation process and uncertainty, though we used Gaussian sample in this work and Gaussian process in our previous work\cite{2021Likelihood} to generate training set. Because the main task in this work is to compare DAE and IMNN, we did not focus on the the simulation model, but our next ongoing work is to build a better model to simulate OHD with deep learning. Secondly, there are other types of autoencoder, such as denoising variational autoencoder (the combination of variational autoencoder \cite{2014Auto} and denoising autoencoder). We chose DAE in this work and our previous work \cite{2021Likelihood} because it can not only learn the robust features but also significantly reduce the noise level. However, it is hard to tell if DAE is the best choice without experiments, so one of our future works is to use the method in this work to compare DAE with other autoencoders.
	
	In the future, we will probably be able to do a better constraint if we can extend our dataset. However, we do not recommend mixing datasets, because it means mixing different errors which are calculated by different methods, we will not necessarily obtain an accurate estimation.
	
	\section*{Acknowledgements}
	We thank the anonymous referee for the comments that helped us greatly improve this paper. The referee reviewed our paper carefully and put forward many useful suggestions. We thank Changzhi Lu, Jin Qin, Jing Niu, Kang Jiao and Tom Charnock for useful discussions and their kind help. This work was supported by the National Science Foundation of China (Grants Nos: 61802428, 11929301), and National Key R\&D Program of China (2017YFA0402600).
	
	\nocite{*}
	
	\bibliographystyle{apsrev4-2}
	\bibliography{apssamp}
	
\end{document}